\documentstyle[12pt]{article}
\makeatletter
\@addtoreset{equation}{section}
\makeatother

\begin{document}
\thispagestyle{empty}
\begin{flushright}
BI-TP 2000/14\\
IRB-TH-5/00\\
hep-ph/0102xxx\\
\end{flushright}
\vspace*{11mm}
\begin{center}
{\Large \bf Pinching phenomenon: Central feature}\\
{\Large \bf in out of equilibrium thermal field theories}\\
\vspace*{11mm}
{\large
 I.~Dadi\'c$^{1,2}$ }
\footnote{E-mail: dadic@faust.irb.hr}
\\[24pt]
$^1$ Ruder Bo\v{s}kovi\'{c} Institute,
Zagreb, Croatia\\
$^2$ Fakult\"at f\"ur Physik, Universit\"at
Bielefeld,
Bielefeld, Germany \\
\vspace*{11mm}
\end{center}
\begin{abstract}
We study out of equilibrium
thermal field theories with switching on the interaction
occurring at finite time. We continue to study a formulation
exploiting the concept of projected functions (PF) and
Wigner transforms of projected functions (WTPF), for which convolution
products between these functions can be achieved in a closed form
without use of the gradient expansion.
Many of the functions,
appearing in the low orders of the perturbation expansion
(bare propagators, one-loop self-energies, retarded and
advanced components of the resummed propagator, ...)
belong to the class of PF or WTPF.
However, WTPF's are completely determined by their
$X_0\rightarrow +\infty$ limit
and, thus, cannot be the carriers of relaxation phenomena.
Furthermore, we observe that the functions capable of
carrying relaxation phenomena (non-WTPF) emerge in the expressions
containing mixed products (i.e., products of retarded and
advanced propagators and self-energies; ill-defined in the usual
formulation with the Keldysh time-path).
In particular, to predict the time dependence of the system,
one has to use equal-time Green functions (particle number, etc.).
These are obtained by
inverse Wigner transform (simple integration over energy in the
case of equal time). The result of this operation is that all
terms originating from WTPF will be constants in time (and
equal to zero in most cases), and only the non-WTPF terms
contribute to time variation. As these are generated in mixed
products, the pinching phenomenon is being promoted from an obstacle
to the central feature of out of equilibrium
thermal field theories.

We analyze the pinching phenomenon in some detail.
In the case of naive pinching (product only of
retarded and advanced components of the bare propagator),
for short times our calculation confirms the existence of
the contributions linear in $X_0$. At very large times
the contribution evolves to the usual pinching singularity.
In Schwinger-Dyson equations the Keldysh component
of self-energy always appears between the powers of
retarded and advanced propagators.
One easily finds that the mathematical
expression corresponding to such a product
is well defined even for multiple
self-energy insertion contributions.
We study the case of single self-energy insertion in more detail.
We obtain the non-WTPF contribution
which generates nontrivial $X_0$ dependence.

In the case of production of a photon from QCD plasma
(finite-lifetime effect) approximate analytic results
from our approach are almost identical
to those obtained by
S. -Y. Wang and D. Boyanovsky, who use the dynamical
renormalization group approach.
\end{abstract}

\section{ Introduction}
Important aspects of modern physics depend very much on our
understanding of nonequilibrium phenomena.

Many years of development of out of equilibrium thermal field theory (TFT)
\cite{schwinger,keldysh}
have resulted in slow but steady progress
\cite
{kadanoff,niemi,danielewicz,cshy,rammer,landsman,calzetta,eboli,remler,mleb,brown,gl,bio}.

For almost equilibrated systems,
at infinite time after switching on the interaction, a number
of results are valid at the lowest order in the gradient expansion
\cite{langreth,henning,bi}:
the cancellation of
collinear \cite{lebellac,niegawacom} and pinching singularities
\cite{as,altherr,bedaque,niegawa,gl2,id}, the extension of
the hard thermal loop (HTL) approximation \cite{pisarski,bp,frenkel}
to out of equilibrium
\cite{carrington,defu}, and applications to heavy-ion collisions
\cite{bdr,bdrs,bdrk}.

For some problems, e.g., heavy-ion collisions, the above limitations
are
undesired. If one wants to consider large deviations from
equilibrium, one should go beyond the gradient
expansion. One cannot wait infinitely long as these systems go
apart after a very short
time, probably without reaching the stage of
equilibrium. In nuclear collisions,
short-time scale features have been studied
in a number of papers
$\cite{grange,tohyama,reinhard,tohyama2,gwr,morawetz}$.

One of the characteristic features of out
of equilibrium TFT is the appearance of
mixed products of retarded and advanced
propagators. In a formulation similar to
equilibrium TFT, which use Keldysh time-path
these terms have led to pinching
singularities (or ill-defined $\delta^2$ expressions).
Many attempts to get rid of these can be
classified as attempts within the zeroth
order in the gradient expansion and as
attempts to go beyond.

The first group of
papers$^{\cite{as,altherr,lebellac,niegawa,id}}$,
although successful in eliminating pinching in some cases
(in the single self-energy insertion contributions
to the Keldysh component of the propagator
in theories like
QED and QCD \cite{id}, but not in multiple
self-energy insertions), do not solve the problem
of pinching in the theories like $\phi^3$, $\phi^4$,
nor in the theory describing the $\rho-\pi$ interaction.

The second group of
papers \cite{bedaque,gl,devega,bv,bvhs,wb,wbvl}
use the finite switching-on time integration
path (see Fig. 1). In these
approaches, the seminal terms (the terms
growing infinitely with $X_0$) appear instead of
pinching .

In our recent paper \cite{hep} we have studied out of equilibrium
thermal field theories with switching on the interaction
occurring at finite time. We observe that many of the functions,
appearing in the low orders of the perturbation expansion,
belong to the class of projected functions (PF in further text) or
Wigner transforms of projected functions (WTPF). These functions
have particularly simple multiplication rules. However, WTPF are
completely determined by their $X_0\rightarrow +\infty$ limit
and, thus, cannot be carriers of relaxation phenomena.
Furthermore, we observe that the functions capable of
carrying relaxation phenomena (non-WTPF) emerge in the expressions
containing mixed products (i.e., products of retarded and
advanced propagators and self-energies). It is important to note that
the non-WTPF contribution emerges even in the case when the
time path is not  "pinched" by two infinitely close poles; it is enough
that the poles (or more complicated singularities)
are situated on the opposite sides
of the integration path. This means that the calculation with resummed
propagators with complex poles (if we manage to have such!) will as well
produce non-WTPF terms and generate relaxation.

In the present paper we develop these ideas further. In particular,
to predict the time dependence of the system,
one has to use equal-time Green functions. These are obtained by
inverse Wigner transform (simple integration over energy in the
case of equal time). The result of this operation is that all
terms originating from WTPF are constants in time (and
equal to zero in most cases), and only the non-WTPF terms
contribute to time variation. As these are generated by mixed
products, the pinching phenomenon is being promoted from an obstacle
to the central feature of out-of-equilibrium
thermal field theories.

In the present paper, after a short recapitulation of
earlier results (Sec. II), we introduce (Sec. II.1) the function
$sign(p_0,\omega_p)$, which is the generalization
of the $sign(p_0)$ function. Especially in the case of particles with spin
one use the adventages of the function $sign(p_0,\omega_p)$
to obtain the expressions "manifestly" retarded or advanced.
In Sec. II.2 we establish the connection between two-point
functions and equal-time functions (number operator, etc.),
and prove that time-dependent
contributions to equal-time functions come solely from
non-WTPF.

In Sec. III we analyze pinching
phenomenon in some detail. We further reduce
the case of naive pinching (product of only a
retarded and an advanced component), to
the problem of pinching between two infinitely close poles.
For short times, our calculation confirms the existence of
the contributions linear in $X_0$. At very long times,
the contribution evolves to the usual pinching singularity.

In Schwinger-Dyson equations (Sec.IV) the Keldysh component
of self-energy always appears between the powers of
retarded and advanced propagators . One easily finds that the mathematical
expression corresponding to such a product
is well defined even for multiple
self-energy insertion contributions (Sec. IV.1). In the
single self-energy insertion case (Sec. IV.2), one obtains two
contributions. The one corresponding to pinching in the Keldysh
time path approach, owing to the "$\epsilon$"-shift becomes
just the usual WTPF consisting of only one type of R/A
components of the propagator and self-energy. The other contribution
is non-FTPF; it generates nontrivial $X_0$ dependence.

In this paper it is important to preserve
strict "$\epsilon$" bookeeping the
importance of which has been known since the discussion on
the proper analytical continuation
between the imaginary time formalism and the R/A approach
\cite{aurenche,eijck,guerin,evans} in the real time formalism at
equilibrium.

Our approach is an alternative to the dynamical
renormalization group (DRG) approach \cite{devega,bv,bvhs}.
Whereas we find the way to work with Feynman diagrams
in the energy-momentum space and do not
use the gradient expansion
(at least in the low orders of the perturbation expansion), in the DRG
approach one relies more on the differential equations with
the gradient expansion as necessary tool.
Nevertheless, the results (in our case, the results of the
research in progress) are sometimes very similar.
For a better understanding,
one should compare the time dependence of specific processes
calculated using both methods.

\section{\bf Out of equilibrium setup}
In our previous paper {\cite{hep}}
we have formulated the approach appropriate for the dynamical situation
arising when the system starts its evolution at finite time (for
simplicity, we take $t_i=0$). In this
formulation, the time integration follows the
finite switching-on time path (see Fig. 1).

To understand the limitations coming from the finite switching-on time,
we start with the two-point function $G(x,y)$.
The quantities $x$ and $y$
are four-vector variables with time components in the range
$0<x_0,y_0<\infty$.
We define the Wigner variables $s$ (relative
space-time, relative variable) and
$X$ (average space-time, slow variable) as usual:
\begin{eqnarray}\label{wvi}
&&X={x+y \over 2},~s=x-y,
\cr
\nonumber\\
&&G(x,y)=G(X+{s \over 2},X-{s \over 2}).
\end{eqnarray}
The lower limit on $x_0$, $y_0$ implies conditions on $X_0$ and $s_0$:
$0<X_0,~-2X_0 < s_0 < 2X_0$.
The two-point function can be expressed in
terms of the Wigner transform (i.e., the Fourier integral with
respect to $s_0, s_i$):
\begin{eqnarray}\label{ft}
G(X+{s \over 2},X-{s \over 2})=(2\pi)^{-4}
\int d^4pe^{-i(p_0s_0-\vec p\vec s)}G(p_0,\vec p;X).
\end{eqnarray}
Here
\begin{eqnarray}\label{ift}
G(p_0,\vec p;X)=\int_{-2X_0}^{2X_0}ds_0\int d^3s
e^{i(p_0s_0-\vec p\vec s)}G(X+{s \over 2},X-{s \over 2}).
\end{eqnarray}

We have found that the low orders in the perturbation expansion are
characterized by the appearance of very special two-point functions,
we call them
projected functions.
Projected functions (truncated, ``mutilated function''
\cite{wiener}) possess the following properties:
the function does not change with $\vec X$ (homogeneity assumption),
it is a function
of ($s_0,\vec s$) within
the interval $-2X_0<s_0<2X_0$ and identical to zero outside the interval:
\begin{eqnarray}\label{prfi}
&&F(X+{s\over 2}, X-{s\over 2})=\Theta(X_0)\Theta(2X_0-s_0)
\Theta(2X_0+s_0)\bar F(s_0,\vec s)
\cr\nonumber\\&&
\bar F(s_0,\vec s)=\lim_{X_0\rightarrow \infty}F(X+{s\over 2},X-{s\over 2}).
\end{eqnarray}
The projected function can be viewed as the projection of the whole function
$F_{\infty}(s_0,\vec s)=F(s_0,\vec s)$ (i.e. the function
defined at $X_0=+\infty$ which uses the
whole $s_0$ axis as a carrier)
to its finite carrier. The projection operator is
$P_{X_0}(s_0)=\Theta(X_0)\Theta(2X_0-s_0)
\Theta(2X_0+s_0)$: $F_{X_0}=P_{X_0}(s_0)F_{\infty}$.
The Wigner transform of the projected function (WTPF)
at the given time $X_0$ may be obtained
using the Wigner transform of the projected function
(WTPF) at the time $X_0=+\infty$.
\begin{eqnarray}\label{iftf}
F_{\infty}(p_0,\vec p)=\int_{-\infty}^{\infty}ds_0\int d^3s
e^{i(p_0s_0-\vec p\vec s)}F(s_0,\vec s),
\end{eqnarray}
and the projection operator $P_{X_0}$
\begin{eqnarray}\label{rftf}
F_{X_0}(p_0,\vec p)=\int_{-\infty}^{\infty}dp'_0
P_{X_0}(p_0,p'_0)F_{\infty}(p'_0,\vec p),
\end{eqnarray}
where
\begin{eqnarray}\label{sftf}
P_{X_0}(p_0,p'_0)={1\over 2\pi}\Theta(X_0)
\int_{-2X_0}^{2X_0}ds_0e^{is_0(p_0-p'_0)}
={1\over \pi}\Theta(X_0){\sin\left( 2X_0(p_0-p'_0)\right)\over p_0-p'_0}
\end{eqnarray}
is the Fourier transform of the projector and
the inverse transform is given by
\begin{eqnarray}\label{isftf}
E^{-is_0p'_0}\Theta(X_0)\Theta(2X_{0}+s_{0})\Theta(2X_{0}-s_{0})
=\int dp_0e^{-is_0p_0}P_{X_0}(p_0,p'_0).
\end{eqnarray}
The
assumption of the homogeneity in space coordinates
excludes any dependence on $\vec X$ and we omit
it as an argument of the function.

It is important to note that
\begin{eqnarray}\label{dsftf}
\lim_{X_0\rightarrow \infty}P_{X_0}(p_0,p'_0)
=\lim_{X_0\rightarrow \infty}{1\over \pi}
{\sin\left( 2X_0(p_0-p'_0)\right)\over p_0-p'_0}
=\delta(p_0-p'_0),
\end{eqnarray}
and
\begin{eqnarray}\label{intP}
\int dp_0P_{X_0}(p_0,p_{01})={1\over 2i\pi}\int_{-\infty}^{\infty}
dp_0\left({e^{2iX_0(p_o-p_{0,1})}-1\over p_o-p_{0,1}}
-{e^{-2iX_0(p_0-p_{0,1})}-1\over p_0-p_{0,1}}\right)=1.
\end{eqnarray}
The last equality is obtained by closing the integration path
in the first term from above and in the second term from below.

Evidently, there is
a hierarchy of the $P_{X_0}$ projectors:
\begin{eqnarray}\label{sss}
P_{X_{0,M}}(p_0,p"_0)=\int dp'_0P_{X_0}(p_0,p'_0)P_{X'_0}(p'_0,p"_0),
~~X_{0,M}=min(X_0,X'_0).
\end{eqnarray}

For further analysis, the analytic properties of the
WTPF in the  $X_0\rightarrow \infty $ limit  as a
function of complex energy are very important.
We define the following properties:
(1) the function of $p_0$ is analytic
above (below) the real axis, (2) the function  goes to zero
as $|p_0|$ approaches infinity in the upper (lower) semiplane.
The choice above (below) and upper (lower) refers to R (A)
components.
It is easy to recognize that the properties (1) and (2) are just the
definition of the retarded (advanced) function.
However, it is nontrivial, and not always true,
that the functions with the R (A) index satisfy them.

Important examples of projected functions satisfying
assumptions (1) and (2) are poles in the energy plane,
retarded, advanced,
and Keldysh components of free propagators, and one-loop self-energies.

The Wigner transform of the convolution product of two-point functions
\begin{eqnarray}\label{pgfi}
C=A*B \Leftrightarrow C(x,y)=\int dz A(x,z)B(z,y)
\end{eqnarray}
is given by the gradient
expansion (note that we have
assumed the homogeneity in space coordinates,
which excludes any dependence on
$\vec X$):
\begin{eqnarray}\label{ge}
C_{X_0}(p_0,\vec p)=e^{-i\diamondsuit}A_{X_0}(p_0,\vec p)
B_{X_0}(p_0,\vec p),~~
\diamondsuit={1\over 2}(\partial_{X_0}^A\partial_{p_0}^B.
-\partial_{p_0}^A\partial_{X_0}^B).
\end{eqnarray}
A much simpler expression is valid for $A$ and $B$ being projected functions:
\begin{eqnarray}\label{iffff}
C_{X_0}(p_0,\vec p)=\int dp_{01}dp_{02}
P_{X_0}(p_0,{p_{01}+p_{02}\over 2})
{1\over 2\pi}{ie^{-iX_0(p_{01}-p_{02}+i\epsilon)}
\over p_{01}-p_{02}+i\epsilon}
A_{\infty}(p_{01},\vec p)B_{\infty}(p_{02},\vec p).
\end{eqnarray}
Under the assumption that $A$ or $B$ satisfy (1) and (2)
($A$ as advanced or $B$ as retarded)
Eq. (\ref{iffff}) can be integrated
even further. We obtain
\begin{eqnarray}\label{ifff4}
C_{X_0}(p_0,\vec p)&=&\int dp'_{0}P_{X_0}(p_0,p'_{0})
A_{\infty}(p'_{0},\vec p)B_{\infty}(p'_{0},\vec p).
\end{eqnarray}
The convolution product of two two-point functions which are WTPF's
and satisfy (1) and (2) is also a WTPF. This product is then expressed
through the projection operator acting on a simple product of two WTPF's
given at $X_0=\infty $.

As expected, in the $X_0=\infty $ limit, Eq.(\ref{ifff4}) becomes a
simple product
\begin{eqnarray}\label{fff4i}
\lim_{X_0\rightarrow \infty}C_{X_0}(p_0,\vec p)=
A_{\infty}(p_{0},\vec p)B_{\infty}(p_{0},\vec p).
\end{eqnarray}

At finite $X_0$, Eq.(\ref{ifff4}) exhibits
a smearing of energy (as much as it is
necessary to preserve the uncertainty relations).

The product of $n$ two-point functions
is obtained by repeating the above procedure:
\begin{eqnarray}\label{ffffn}
&&C_{X_0}(p_0,\vec p)=\int \prod_{j=1}^{n-1}(dp_{0,j})dp_{0,n}
P_{X_0}(p_0,(p_{0,1}+p_{0,n})/2)
\cr
\nonumber\\
&&
\times\prod_{j=1}^{n-1}\left(A_{j,\infty}(p_{0,j},\vec p)
{1\over 2\pi}{i \over p_{0,j}-p_{0,j+1}+i\epsilon}\right)
e^{-iX_0(p_{0,1}-p_{0,n}+i(n-1)\epsilon)}
A_{n,\infty}(p_{0,n},\vec p).
\end{eqnarray}
We note here: the condition that the intermediate products should also
be projected functions requires that at least n-1 of the functions
in the product should satisfy assumptions (1) and (2) (the retarded should
be on the right-hand side and the advanced on the left-hand side, and
the function that eventually does
not satisfy (1) and (2) should be inbetween). However,
this is not the order in which
the components appear in the Schwinger-Dyson equation.

Then  one can perform all integrations except
one to obtain
\begin{eqnarray}\label{fff4n}
C_{X_0}(p_0,\vec p)=\int dp_{0,1}P_{X_0}(p_0,p_{0,1})
\prod_{j=1}^{n}A_{j,\infty}(p_{0,1},\vec p).
\end{eqnarray}
\subsection{\bf Function $sign(p_0,\omega_p)$, propagator and self-energy}
For the retarded (advanced) and Keldysh components of the
bare propagator one obtains
(owing to the relation (\ref{rftf})
it is enough to give their form at infinite time $X_0={+\infty}$):
\begin{eqnarray}\label{fnonigrkr}
G_{R(A),\infty}(p)=(-G_{1,1}+G_{1,2(2,1)
})_{\infty}(p)={-i\over p^2-m^2\pm 2i\epsilon p_0}.
\end{eqnarray}
\begin{eqnarray}\label{fnoigrkr1}
G_{K,\infty}(p)=(G_{1,1}+G_{2,2})_{\infty}(p)=
2\pi[1+2f(\omega_p)]\delta(p^2-m^2).
\end{eqnarray}
Our "$_{\infty}$"
components coincide with the usual Keldysh-integration-path
propagators. These expressions are easily
generalized to the case of other spin and
statistic assignment.

The Keldysh component as a function of $p_0$ does not satisfy
assumptions (1) and (2), however,
we can decompose it into the sum
of functions which satisfy them either as retarded or as advanced
functions.
This trick we repeat below in the case of $\Sigma_K$.
To do so we start with an identity:
\begin{eqnarray}\label{ident1}
\delta(x-y)={i\over 2\pi}\gamma({x\over y})
[{1\over x-y+i\epsilon}-{1\over x-y-i\epsilon}]+{\cal O}(\epsilon^2),
\end{eqnarray}
where $\gamma(1)=1$, otherwise, $\gamma(x/y)$ is only weakly constrained:
it should be analytic around $x/y=1$. Equation (\ref{ident1}) is
used to generate the following identity:
\begin{eqnarray}\label{ident2}
\delta(p_0^2-\omega_p^2)={i\over 2\pi}sign({p_o,\omega_p})
[{1\over p_0^2-\omega_p^2+2i\epsilon p_0}
-{1\over p_0^2-\omega_p^2-2i\epsilon p_0}]+{\cal O}(\epsilon^2).
\end{eqnarray}
In the above identity we have substituted the
usual $sign(p_0)$ function by a new (user friendly)
function $sign(p_0,\omega_p)$, which
we specify as an alternative between
\begin{eqnarray}\label{ssf}
sign(p_0,\omega_p)=sign(p_o),
{p_0\over \omega_p},
{\omega_p\over p_0},
\left({p_0\over \omega_p}\right)^{3},
\left({\omega_p\over p_0}\right)^{3},...
\end{eqnarray}
Evidently, the function $sign(p_0,\omega_p)$ at $p_0=\pm\omega_p$
for all offered possibilities
reduces to $sign(p_0)$ and the identity is valid.
The choice of the appropriate form of $sign(p_0,\omega_p)$ should guarantee
that in the perturbative expansion integrals over $p_0$ converge
(in such a way that two terms in Eq. (\ref{fnonigrkr1}),
$G_{K,R}$ and $G_{K,A}$, could be
treated separately)
at $|p_0|=\infty$ and no additional singularities appear at finite $p_0$
(especially not at $|p_0|=0$). This choice might be different
for different terms.
The difference between any two choices (when multiplied by
$\delta(p_0^2-\omega_p^2)$) is $\cal{O}$$(\epsilon^2)$.
In the absence of pathology, this difference vanishes in the
$\epsilon\rightarrow 0$ limit.
The usual $sign(p_0)$
(first, not a recommended {\cite{id}} choice),
owing to its nonanalytic nature, has prevented
the use of Cauchy integrals in the expressions containing
$G_{K,R,\infty}$. The choice ${p_0\over \omega_p}$ is a default choice,
one uses it if the integrals converge. The choice ${\omega_p\over p_0}$
is useful; with respect to the default choice,
it reduces the power of $p_0$ by two units; if there is
a factor $p_0$ in the integrand, this choice will not produce
extra singularities at $p_0=0$. Similarly one can decide on the use of
other choices.
Having made proper choices, in the loop
integration one can integrate over $p_0$ as first. This will
result in manifestly retarded (advanced) functions.
The $\epsilon$ parameter, which regulates these expressions,
should be kept uniformly finite
during the calculations, and the limit $\epsilon \rightarrow 0$
should be taken last of
all \cite{landsman}. This specially means that
$\lim_{X_0\rightarrow\infty}\exp(-X_0\epsilon)=0$
and the terms containing this factor vanish in the
$X_0\rightarrow\infty$ limit.
Now we can write
\begin{eqnarray}\label{fnonigrkr1}
&&G_{K,\infty}(p)=
-G_{K,R,\infty}(p)+G_{K,A,\infty}(p),
\cr\nonumber\\
&&G_{K,R(A),\infty}(p)=(1+2f(\omega_p))
sign(p_0,\omega_p)G_{R(A),\infty}(p).
\end{eqnarray}

To discuss the amputated one-loop self-energy, we start with
(spin and internal
symmetry indices are suppressed)
\begin{eqnarray}\label{se}
&&\Sigma(x,y)=ig^2S(x,y)D(x,y),
\cr\nonumber\\&&
\Sigma_{R(A)}(p)=-(\Sigma_{1,1}+\Sigma_{1,2(2,1)})(p),
\cr\nonumber\\
&&\Sigma_{R(A),\infty }(p)=\pm{ig^2\over 2}\int {d^4k\over (2\pi)^4}
[h(k_0,\omega_k)+h(p_0-k_0,\omega_{p-k})]
D_{R(A),\infty}(k)S_{R(A),\infty}(p-k)F,\cr\nonumber\\
&&\Sigma_K(p)=(\Sigma_{11}+\Sigma_{22})(p)
=-\Sigma_{K,R}(p)+\Sigma_{K,A}(p)
\cr\nonumber\\
&&\Sigma_{K,R(A),\infty}(p)=\mp{ig^2\over
2}\int {d^4k
\over (2\pi)^4}
[1+h(k_0,\omega_k)h(p_0-k_0,\omega_{p-k})]
\cr\nonumber\\&&
D_{R(A),\infty}(k)S_{R(A),\infty}(p-k)F,
\end{eqnarray}
where $D$ and $S$ are bare scalar propagators,
$h(k_0,\omega_k)=-sign(k_0,\omega_k)1+2f(\omega_p)]$
and the factor $F=F(k_0,|\vec k|,p_0,|\vec p|,\vec k\vec p,...)$
includes the information about spin and internal degrees of freedom
($F=1$ if all particles are scalars).

In the calculation of R, A, and K components of self-energy we use general
expressions  given by Eqs. (II.23)-(II.25) of Ref. \cite{id}.
For particles with spin  it is the appropriate choice of
$sign(k_0,\omega_k)$ that makes the integral
over $k_0$ convergent term-by-term.
Then the integrals over $D_RS_A$ and $D_AS_R$ vanish; we are
thus left with the pure RR(AA) cotribution to the R(A) component.

General analytic properties of the expressions
of the type (\ref{se}) are well known: there
are discontinuities (cuts) along the real axis (or, better to say,
displaced from the real axis by $-i\epsilon$ ($+i\epsilon$) for
the retarded (advanced) component), starting at thresholds
for various real processes.
\subsection{\bf Equal-time two-point functions}
To define single-particle observables one has to study reduction
of two-point functions to equal time ($x_0=y_0=t$ or
$X_0=t$, $s_0=0$) \cite{kadanoff}.
These can be obtained by inverse Wigner transform as
\begin{eqnarray}\label{etc}
G(t,0,\vec p)={1\over 2\pi}\int dp_{0}G_{X_0=t}(p_{0},\vec p).
\end{eqnarray}

As an example of equal-time two-point function one can study the number
operator.
To define it, we start with the
Keldysh component of the propagator:
\begin{eqnarray}\label{3wx}
G_K(x,y)=G_{1,2}(x,y)+G_{2,1}(x,y)=
<\phi(x)\phi(y)+\phi(y)\phi(x)>,
\end{eqnarray}
\begin{eqnarray}\label{3wk}
G_K(X_0,s_0,\vec p)
=\int d^3se^{-i\vec p \vec s}G_K(X+{s\over 2},X-{s\over 2})
=(2\pi)^{-1} \int dp_0e^{ip_0s_0}G_{X_0,K}(p_0,\vec p).
\end{eqnarray}
At $x_0=y_0=t$ (i.e., $s_0=0$; $X_0=t$) one finds
(under the usual assumption that $<aa>$ and $<a^+a^+>$ terms
vanish), the relation between the number operator and the Keldysh
component of the propagator is
\begin{eqnarray}\label{0t}
<2N_{\vec p}(t)+1>=\omega_{p}G_K(t,0,\vec p)
={\omega_{p}\over 2\pi}\int dp_0G_{t,K}(p_0,\vec p).
\end{eqnarray}
Further single-particle observables are generated with the help of
$<N_{t}>$. In the case of bare fields, one
obtains as expected:
\begin{eqnarray}\label{fnonigrkrd}
&&<2N^0_{\vec p}+1>={\omega_p\over 2\pi}\int dp_0 G^0_{X_0,K}(p)
={\omega_p\over 2\pi}\int dp_0dp'_0P_{X_0}(p_0,p'_0)G^0_K(p'_0,\vec p)
\cr\nonumber\\&&
={\omega_p\over 2\pi}\int dp'_0 G^0_K(p'_0,\vec p)
=-Im(\int dp'_0{p'_0\over\pi}
{1+2f(\omega_p)\over p'^2_0-\omega_p^2+ 2i\epsilon p'_0}),
\cr\nonumber\\&&
=1+2f(\omega_p).
\end{eqnarray}
The time independence of the right-hand side of
relation (\ref{fnonigrkrd}) is a special case of a more general feature.

An equal-time two-point function coming from a retarded WTPF
may be obtained with the help of Eqs. (\ref{etc}), (\ref{rftf}),
and(\ref{intP}) as (to avoid problems with $\Theta$'s, we understand here
that setting $s_0=0$ is achieved by the limiting procedure
$\lim_{s_0\rightarrow +0}$):
\begin{eqnarray}\label{iccc}
&&G_R(t,0,\vec p)={1\over 2\pi}\int dp_0G_{X_0,R}(p_0,\vec p)
\cr\nonumber\\&&
={1\over 2\pi}\int dp_0\int dp_{01}P_{X_0}(p_0,p_{01})
G_{\infty,R}(p_{01},\vec p)
\cr\nonumber\\&&
={1\over 2\pi}\int dp_{01}G_{\infty,R}(p_{01},\vec p)
=const(\vec p).
\end{eqnarray}
The integral over the WTPF $G_{X_0,R}$ does not change with time
 as the right-hand side
refers to $\infty$ and not to $X_0$.It is even vanishing for
expressions containing two or more bare retarded propagators in
the product, as one can easily see
by closing the path of integration over $p_{0,1}$ in
Eq. (\ref{iccc}) from above.
One obtains the same result for the advanced function by closing
the integration path from below.

This is a very important result, but is by no means surprising:
the projected function is completely
determined by its form at $X_0=+\infty$ If it were to describe
irreversible processes, it would violate causality. Now we may conclude
that one really needs non-WTPFs to describe the time dependence of
single-particle observables.
These will emerge as a by-product of pinching.
\section{ Examples of pinching }
\subsection{Naive pinching with retarded and
advanced propagators}
The naive pinching singularity is represented
by (at $X
_0=\infty $)
\begin{eqnarray}\label{naive}
G_{pinch}=G_{R}*G_{A},
\end{eqnarray}
where $G_{R(A)}$ is given by (\ref{fnonigrkr}).
One can decompose $G_{R(A),\infty}$ into the
sum of two poles
\begin{eqnarray}\label{naive2}
G_{R(A),\infty}(p)={-i\over
2\omega_p}\left(
{1\over p_0\pm i\epsilon-\omega_p}
-{1\over p_0\pm i\epsilon+\omega_p}\right),
\end{eqnarray}
so it is enough to study pinching between two
infinitely close poles.
\subsection{ Pinching between two infinitely close poles}
We assume the contribution of the pole infinitely close to the real axis:
\begin{eqnarray}\label{p0}
{\cal G}_{R(A),\infty,pole}(p_0)={1\over p_0-\bar p_0\pm i\mu},
\end{eqnarray}
where $\bar p_0$ is real and $\mu>\epsilon/2$.
It satisfies assumptions (1) and (2)
with the $+$ sign as retarded function,
and with the $-$ sign as advanced functions.

The product $G_{R,pole}*G_{A,pole}$ is easily obtained by
substituting $G_{R(A),\infty,pole}$ into (\ref{iffff}).
We choose new variables
$P_0=(p_{01}+p_{02})/2$ and $\Delta_0=p_{01}-p_{02}$,
and integrate over $\Delta_0$
(care is necessary as $\epsilon $ and $\mu $ are both infinitely
small quantities) to obtain
\begin{eqnarray}\label{raint}
&&C_{X_0}(p_0)=\int dP_{0}P_{X_0}(p_0,P_{0})
{1\over P_0-\bar p_0+i\mu-i\epsilon/2}{1\over P_0-\bar p_0-i\mu+i\epsilon/2}
\cr
\nonumber\\
&&+\int dP_{0}P_{X_0}(p_0,P_{0}){1\over \bar p_0-p_0}
\left({e^{2iX_0(P_0-\bar p_0+i\mu-i\epsilon/2)}\over
2(P_0-\bar p_0+i\mu)-i\epsilon}
+{e^{-2iX_0(P_0-\bar p_0-i\mu+i\epsilon/2)}\over
2(P_0-\bar p_0-i\mu)+i\epsilon}\right).
\end{eqnarray}
The first term is the projected function; in the
$X_0\rightarrow \infty $, limit
it becomes a usual example of pinching. The second term
consists of two non-WTPF pieces.

Further integration gives
(after introducing $\rho=\mu-\epsilon/2$)
\begin{eqnarray}\label{rapols}
C_{X_0}(p_0)={1
-e^{-2X_0\rho}\cos2X_0(p_0-\bar p_0)
\over (p_0-\bar p_0+i\rho)(p_0-\bar p_0-i\rho)}
+{\rho e^{-2X_0\rho}\sin2X_0(p_0-\bar p_0)
\over (p_0-\bar p_0+i\rho)(p_0-\bar p_0-i\rho)(p_0-\bar p_0)}.
\end{eqnarray}
Expression (\ref{rapols}) can be studied at different times.
At very very large, but finite time (i.e., such that
$\kappa/\rho\exp (-2X_0\rho)<<1$,
where $\kappa $ is a typical energy scale of the problem) one can ignore
the $exp(-2X_0\rho)$ terms and obtains the usual Keldysh-path pinching.
Needless to say, as $\rho $ is arbitrarily small, the time should
be "arbitrarily very very" large.

Ignoring the intermediate scales, we come to the finite-time scale.
At this scale
$X_0\rho<<1$ and the exponential can be substituted by "$1$".
For large times ($X_0>>\kappa^{-1}$), one can approximate {\cite{bvhs}}
\begin{eqnarray}\label{intapr}
&&{\sin2X_0(p_0-\bar p_0)\over p_0-\bar p_0}\approx \pi\delta(p_0-\bar p_0),
\cr\nonumber\\
&&{\rho\sin2X_0(p_0-\bar p_0)\over (p_0-\bar p_0+i\rho)(p_0-\bar p_0-i\rho)
(p_0-\bar p_0)}\approx 2\pi X_0\delta(p_0-\bar p_0),
\cr\nonumber\\
&&\int_{-\infty}^{\infty}dp_0 f(p_0)
{\sin^2X_0(p_0-\bar p_0)\over (p_0-\bar p_0)^2}\approx
\pi X_0f(\bar p_0)+{{\cal P}\over 2}
\int_{-\infty}^{\infty}dp_0 {f(p_0)-f(\bar p_0)\over(p_0-\bar p_0)^2}.
\end{eqnarray}
Finally, one obtains
\begin{eqnarray}\label{intpapr1}
\int C_{X_0}(p_0)f(p_0)dp_0\approx
4\pi X_0f(\bar p_0)+
\int dp_0{\cal P}{f(p_0)-f(\bar p_0)\over(p_0-\bar p_0)^2}.
\end{eqnarray}
In this expression there is a term proportional to
$X_0\delta(p_0-\bar p_0)$ (seminal
term according to some authors).

As expected, naive pinching at finite times gives
contributions proportional to $X_0$;
at "very very large" times it develops usual pinching singularities.
\section{Elimination of pinching in Schwinger-Dyson equations }

We write the Schwinger-Dyson equations in the form
\begin{eqnarray}\label{sde}
&&{\cal G}_R=G_R+iG_R*\Sigma_R*{\cal G}_R,~~
{\cal G}_A=G_A+iG_A*\Sigma_A*{\cal G}_A,
\cr
\nonumber\\
&&{\cal G}_K=iG_R*\Sigma_K*{\cal G}_A
+iG_K*\Sigma_A*{\cal G}_A+iG_R*\Sigma_R*{\cal G}_K.
\end{eqnarray}
We can expand Eqs. (\ref{sde}) to obtain
($\Sigma_K=-\Sigma_{K,R}+\Sigma_{K,A}$):
\begin{eqnarray}\label{sdse}
&&{\cal G}_R=\sum_{n=0}^{\infty} (G_R*i\Sigma_R*)^nG_R,\cr
\nonumber\\
&&{\cal G}_K=\sum_{n=0}^{\infty}G_{K,n},\cr
\nonumber\\
&&G_{K,n}=-(G_R*i\Sigma_R*)^nhG_R+hG_A(*i\Sigma_A*G_A)^n\cr
\nonumber\\&&
+\sum_{p=0}^{n-1}G_R(i\Sigma_R*G_R)^p*(-\bar\Sigma_{K,R}+\bar\Sigma_{K,A})
(i\Sigma_A*G_A)^{n-p-1},
\end{eqnarray}
where $\bar \Sigma_{K,R(A)}=h\Sigma_{R(A)}+\Sigma_{K,R(A)}$.
Equations (\ref{sdse})
are the forms in which pinching appears in out  of equilibrium
thermal field theories.

In fact, expression (\ref{sdse}) is, term by term,
free of pinching. To
see this, one chooses a typical term
containing $\bar\Sigma_{K,R}$ (the terms containing
$\bar\Sigma_{K,A}$ are then obtainable by complex
conjugation). For fixed $n$ and $m=n-p-1$, one can
perform all integrations between either
$RR$ or $AA$ factors (note the bookkeeping of $\epsilon$'s)
\begin{eqnarray}\label{krnm}
&&
{\cal G}_{K,R,n,m}=(G_R*i\Sigma_R*)^nG_R*
(-i\bar\Sigma_{K,R})*(G_A*i\Sigma_A*)^mG_A,
\cr\nonumber\\
&&{\cal G}_{X_0,K,R,n,m}=\int dp_{0,1}dq_{0,1}
P_{X_0}(p_0,{p_{0,1}+q_{0,1}\over 2})
\cr\nonumber\\
&&\prod_{j=0}^{n-1}\left(G_{\infty,R}(p_{0,1}+i2j\epsilon)
i\Sigma_{\infty,R}(p_{0,1}+i(2j+1)\epsilon)\right)
G_{\infty,R}(p_{0,1}+i2n\epsilon)
\cr\nonumber\\
&&(-i\bar\Sigma_{\infty,K,R}(p_{0,1}+i(2n+1)\epsilon))
{i\over 2\pi}{e^{-iX_0(p_{01}-q_{01}+i2(m+n+1)\epsilon)}
\over p_{01}-q_{01}+i2(m+n+1)\epsilon}G_{\infty,A}(q_{0,1}-i2m\epsilon)
\cr\nonumber\\
&&
\prod_{l=0}^{m-1}\left(
i\Sigma_{\infty,A}(q_{0,1}-i(2m-2l-1)\epsilon)
G_{\infty,A}(q_{0,1}-i2(m-l-1)\epsilon)
\right).
\end{eqnarray}
Owing to the poles of $G_R$ and the cuts of $\Sigma_R$ and
$\Sigma_{K,R}$ below the real axis, and the divergence
of the $e^{-iX_0p_{0,1}}$ factor when $Imp_{0,1} \rightarrow +\infty $,
the integral over $p_{0,1}$
( for similar reasons, also the integral
over $q_{0,1}$) cannot be
evaluated analytically. To find the
analytical properties of ${\cal G}_{K,R,n,m}$, we
have to study integrals of the type
\begin{eqnarray}\label{ikrnm}
{\cal I}_{X_0}(p_{0,1},\epsilon,r)=\int_{-\infty}^{+\infty}dq_{0,1}
{e^{iX_0q_{0,1}}
\over p_{0,1}-q_{0i,1}+ir\epsilon}F_{\infty,A}(q_{0,1}),
\end{eqnarray}
where $r>1$. The function F possesses the singularities only within
the strip $\epsilon<Im q_{0,1}<r\epsilon$
and vanishes when $|q_{0,1}|\rightarrow \infty$ outside
the strip.Thus it satisfies assumptions (1) and (2)
in the lower semiplane.It is easy to see that for $Im p_{0,1}>0$
or $Im p_{0,1}<-c-r\epsilon$, where
$c>0$ is a small finite number, the
integration path can be shifted down away from the strip with
singularities. The integration over
$q_{0,1}$ is
regular even in the $\epsilon\rightarrow 0$ limit, and
${\cal I}(p_{0,1})$ represents the function
analytical in $p_{0,1}$ for $p_{0,1}$ outside the strip
$-c-r\epsilon<Im p_{0,1}<0$ and satisfying
$|Im p_{0,1}|<\infty$.

Also in the integration over $p_{0,1}$ along the real axis, all
"nearby" singularities are confined within the strip below the real axis
and one can move
the integration path for the $p_{0,1}$ integration uphill to obtain regular
integrals. Thus there is no pinching in the $p_{0,1}$ integration.

Owing to the fact that the factor
$P_{X_0}(p_0,{p_{0,1}+q_{0,1}\over
2})={1\over i\pi}{e^{iX_0(2p_0-p_{0,1}-q_{0,1})}-e^{-iX_0(2p_0-p_{0,1}-q_{0,1})}
\over 2p_0-p_{0,1}-q_{0,1}}$ is regular at
$2p_0-p_{0,1}-q_{0,1}=0$ and for all $p_0$, $p_{0,1}$,
and $q_{0,1}$ satisfying $|Im p_0|$,$|Im p_{0,1}|$,
and $|Im q_{0,1}|<\infty$, its presence in (\ref{krnm}) will not change our
conclusion that ${\cal G}_{X_0,R,n,m}$ as represented in (\ref{krnm}) is
free from pinching.

Thus we have shown that pinching is absent from the contributions to
${\cal G}_K$ with an arbitrary number of
self-energy insertions.

In the single self-energy insertion approximation, one can perform the
proof in more detail.
\subsection{Elimination of pinching in the
single-self-energy insertion approximation}
The  single-self-energy-insertion approximation to the Keldysh component
of the propagator is expressed as \cite{id} (we treat only the scalar case,
superscript "$0$" bare,
superscript "$1$" one-loop contribution)
\begin{eqnarray}\label{sdsol2p}
&&G_{K}=G^1_{Kp,R}+G^1_{Kp,A}+G^0_{Kr}+G^1_{Kr}+...,
\cr\nonumber\\
&&G^1_{Kp,R}=-iG_R*\bar\Sigma_{K,R}*G_A,~~G^1_{Kp,A}=iG_R*\bar\Sigma_{K,A}*G_A,
\cr\nonumber\\
&&G^0_{Kr}+G^1_{Kr}=h(G_R-G_A)+iG_R*h\Sigma_R*G_R
\cr\nonumber\\
&&-iG_A*h\Sigma_A*G_A.
\end{eqnarray}
In expression (\ref{sdsol2p}), $G^1_{Kp,R}$ and $G^1_{Kp,A}$ are
 potentially ill-defined, while $G^0_{Kr}$ and $G^1_{Kr}$ are
 explicitly free from pinching.

To see what happens in full detail, we start with the contribution
containing $\bar\Sigma_{K,R}$ (we do not indicate
explicitly the dependence on $\vec p$ on the
right-hand sides of the following equations):
\begin{eqnarray}\label{sdsol2R}
&&G^1_{Kp,R}=-iG_R*\bar\Sigma_{K,R}*G_A,
\cr\nonumber\\
&&G^1_{X_0,Kp,R}(p_o,\vec p)=-i\int dp_{01}dp_{02}dp_{03}
P_{X_0}(p_0,{p_{01}+p_{03}\over 2})
G_R(p_{01}){i\over 2\pi}
{e^{-iX_0(p_{01}-p_{02}+i\epsilon)}\over p_{01}-p_{02}+i\epsilon}
\cr\nonumber\\
&&\bar\Sigma_{\infty,K,R}(p_{02})
{i\over 2\pi}
{e^{-iX_0(p_{02}-p_{03}+i\epsilon)}\over p_{02}-p_{03}+i\epsilon}
G_A(p_{03}),
\end{eqnarray}
Here we can integrate over $p_{02}$ by closing the integration
path from above. The only singularity closed in the integration
path is situated at $p_{01}+i\epsilon $.
The result is (note the care for $\epsilon $'s):
\begin{eqnarray}\label{sdsol2R1}
&&G^1_{X_0,Kp,R}(p_0,\vec p)=
-i\int dp_{01}dp_{03}P_{X_0}(p_0,{p_{01}+p_{03}\over 2})
G_R(p_{01})
\cr\nonumber\\
&&\bar\Sigma_{\infty,K,R}(p_{01}+i\epsilon){i\over 2\pi}
{e^{-iX_0(p_{01}-p_{03}+2i\epsilon)}\over p_{01}-p_{03}+2i\epsilon}
G_A(p_{03}),
\end{eqnarray}
Now one can integrate over $p_{03}$ by closing the integration
path from above. The  singularities closed within the
path are situated at $p_{01}+2i\epsilon$ and at
$\pm\omega_p+i\epsilon $.
\begin{eqnarray}\label{sdsol2R2}
&&G^1_{X_0,Kp,R}(p_0,\vec p)=-i\int dp_{01}P_{X_0}(p_0,p_{01})
G_R(p_{01})\bar\Sigma_{\infty,K,R}(p_{01}+i\epsilon)G_A(p_{01}+2i\epsilon)
\cr\nonumber\\
&&+{1\over 2\omega_p}\int dp_{01}
G_R(p_{01})\bar\Sigma_{\infty,K,R}(p_{01}+i\epsilon)
\sum_{\lambda=-1}^1\lambda P_{X_0}(p_0,{p_{01}+\lambda \omega_p\over 2})
{e^{-iX_0(p_{01}-\lambda \omega_p+i\epsilon)}\over
p_{01}-\lambda \omega_p+i\epsilon}.
\end{eqnarray}
By inspecting the definitions of $G_R$ and $G_A$ in (\ref{fnonigrkr}),
one observes that $G_A(p_{01}+2i\epsilon)=G_R(p_{01})$, so that
all functions appearing in (\ref{sdsol2R2})
are retarded. There is no pinching, but we have obtained functions
depending directly on time $X_0$, i.e.,
non-WTPF functions, which one cannot
convolute further in an elegant way we have used here.

One can do the same with the term containing $\bar\Sigma_{K,A}$, but now one
has to integrate over $p_{02}$ by closing the path from below, and
over $p_{01}$ again closing path from below, the result is
(now one needs $G_R(p_{03}-2i\epsilon)=G_A(p_{03})$)

\begin{eqnarray}\label{sdsol2A2}
&&G^1_{X_0,Kp,A}(p_0,\vec p)=i\int dp_{03}P_{X_0}(p_0,p_{03})
G_R(p_{03}-2i\epsilon)\bar\Sigma_{\infty,K,A}(p_{01}+i\epsilon)G_A(p_{01})
\cr\nonumber\\
&&-{1\over 2\omega_p}\int dp_{03}
\sum_{\lambda=-1}^1\lambda P_{X_0}(p_0,{p_{03}+\lambda \omega_p\over 2})
{e^{iX_0(p_{03}-\lambda \omega_p-i\epsilon)}\over
p_{03}-\lambda \omega_p-i\epsilon}
\bar\Sigma_{\infty,K,A}(p_{03}-i\epsilon)G_A(p_{03}).
\end{eqnarray}
Now we
add $G_{Kr}$ to (\ref{sdsol2R2}) and (\ref{sdsol2A2})
and obtain (we can ignore
"the surplus of $\epsilon$" in $\bar\Sigma_{K,R(A)}$)
\begin{eqnarray}\label{sdsol2RA2}
&&G^1_{X_0,K}(p_0,\vec p)=2Im\left(\int dp_{01}P_{X_0}(p_0,p_{01})
G_R(p_{01},\vec p)\Sigma_{\infty,K,R}(p_{01},\vec p)
G_R(p_{01},\vec p)\right.
\cr\nonumber\\
&&\left.+i{1\over 2\omega_p}\int dp_{01}
G_R(p_{01},\vec p)\bar\Sigma_{\infty,K,R}(p_{01},\vec p)
\right.\cr\nonumber\\&&\left.
\sum_{\lambda=-1}^1\lambda P_{X_0}(p_0,{p_{01}+\lambda \omega_p\over 2})
{e^{-iX_0(p_{01}-\lambda \omega_p-i\epsilon)}\over
p_{01}-\lambda \omega_p+i\epsilon}\right).
\end{eqnarray}
This expression is a function of two variables,
$p_0$ and $\vec p$. It is the generalization of the
usual mass shell condition. For fixed $\vec p$,
it offers information about the shape of the
distribution of particles as a function of time.
If we are not interested in the shape of the distribution,
we can integrate over $p_0$. The result is a one-loop
contribution to the number operator. It
tells us about the time dependence of
the occupancy of a given
set of particle states characterized by fixed
$\vec p$. As the first term in Eq. (\ref{sdsol2RA2})
is a retarded function,  it vanishes after integration.
Thus the projected function does not change the integrated
distribution function! It only redistributes
the given contribution within the shape. The function is
symmetric under the change $p_0\rightarrow -p_0$;
thus division by $2$ is equivalent to the projection
to positive frequencies.

Yhe second term can be rearranged to obtain ($X_0=t$)
\begin{eqnarray}\label{intp02RA2}
&&<2N_{\vec p}(t)+1>=<2N^0_{\vec p}(t)+1>+<2N^1_{\vec p}(t)>+...
\cr\nonumber\\
&&={\omega\over 2\pi}\int dp_0G^0_{X_0,K}(p_0,\vec p)
+{\omega\over 2\pi}\int dp_0G^1_{X_0,K}(p_0,\vec p)
\cr\nonumber\\&&
=1+2f(\omega_p)+{\omega\over \pi}Im\left(\int dp_{01}
G_R(p_{01})\bar\Sigma_{\infty,K,R}(p_{01})G_R(p_{01})\right.
\cr\nonumber\\
&&\left.
[1-e^{-iX_0(p_{01}+i\epsilon)}
(\cos X_0\omega_p+i{p_{0,1}\over \omega_p}\sin X_0\omega_p)]\right).
\end{eqnarray}
The term proportional to 1 in braces is added for convenience;
it vanishes upon integration over $p_{01}$ in the upper hemisphere.
The fact that WTPF do not contribute to Eq.
(\ref{intp02RA2}) throws a new light on our
approach: pinchlike contributions ( i.e., those
containing convolution products of both
retarded and advanced components) are
necessary to obtain nontrivial time dependence.
As this fact will reappear in other
expressions (even the calculation of retarded and
advanced components from two-loop or more
complicated Feynman diagrams)
we may conclude that, indeed, pinchlike
expressions represent "the body of evidence"
that very important information is left
"ill-defined" in the formulation using
the Keldysh time path.

To understand the meaning of (\ref{intp02RA2})
we have to compare it with Eq. (10) from \cite{wb}
(see also \cite{wbvl}) for
"enhanced photon production from quark-gluon plasma -
finite-lifetime effect":
\begin{eqnarray}\label{intwb}
<N^1_{\vec p}(t)>=<N_{\vec p}(0)>+
{2\over \pi(2\pi)^3}
\int_{-\infty}^{\infty} dp_{01}R(p_{01})
{1-\cos[(p_{01}-\omega_p)t]\over \pi (p_{01}-\omega_p)^2}.
\end{eqnarray}
The differences are: 1) In this paper we treat
only the scalar case. To adopt it for vector photons and spinor quarks,
we have to substitute $\Sigma(p_0)\rightarrow 2\Sigma^T(p_0)$ (T for
projection of $\Sigma$ to its transverse part, factor
2 for two transverse degrees of freedom).
2) Wang and Boyanovsky use $R(p_0)=-Im\bar\Sigma_<(p_0)$,
while we prefer the Keldysh component $\bar\Sigma_{K,R(A)}(p_0)$.
The difference is a WTPF, so it does not contribute. However, there is
the  term with $Re\bar\Sigma_{K,R(A)}(p_0)$, which is not present
in (\ref{intwb}). It gives rise to extra oscilations.
3) We have used $-\omega_p^2G_R(p_o,\vec p)^{2}$,
while in (\ref{intwb}) one finds
$(p_{01}-\omega_p)^{-2}$. Owing to this difference
their integrand is no longer symmetric under
the change $p_{01}\rightarrow -p_{01}$. This will be more
important for contributions from larger $|p_{01}-\omega_p|$.
This difference seems to come from the approximations intrinsic to
the dynamical renormalisation group approach.
4) Nevertheless an approximate analytical calculation
using HTL self-energies as input to photon-quark-antiquark
analog (\ref{intp02RA2}) and to (\ref{intwb}) \cite{wb} gives almost
identical results.
\section{Conclusions}
We have studied out of equilibrium
thermal field theories with switching on the interaction
occurring at finite time. We have continued to study formulation
exploiting the concept of projected function (PF) and
Wigner transform of projected function (WTPF),for which convolution
products between these functions can be achieved in a closed form
without use of the gradient expansion.
Many of the functions,
appearing in the low orders of the perturbation expansion
(bare propagators, one-loop self-energies, retarded and
advanced components of resummed propagator, ...)
belong to the class of PF or WTPF.
However, WTPF's are completely determined by their
$X_0\rightarrow +\infty$ limit
and, thus, cannot be the carriers of relaxation phenomena.
Furthermore, we have observed that the functions capable of
carrying relaxation phenomena (non-WTPF) emerge in the expressions
containing mixed products (i.e., the products of retarded and
advanced propagators and self-energies; ill-defined in the usual
formulation with the Keldysh time-path).
In particular, to predict the time dependence of the system,
one has to use equal-time Green functions (particle number, etc.).
These are obtained by
inverse Wigner transform (simple integration over energy in the
case of equal time). The result of this operation is that all
terms originating from WTPF's will be constants in time (and
equal to zero in most cases), and only non-WTPF terms
contribute to time variation. As these are generated in mixed
products, the pinching phenomenon is being promoted from an obstacle
to the central feature of out of equilibrium
thermal field theories.

We have analyzed pinching phenomenon in some details.
A general feature here is that in the expressions
containing pinching in the Keldysh time-path
formulation, simple products of retarded
and advanced components become double
integrals of corresponding quantities.

In the case of naive pinching (product only of
retarded and andvanced component),
at short times, our calculation confirms the existence of
contributions linear in $X_0$. At very large times
the contribution evolves to the usual pinching singularity.
In this case pinching singularity appears
as an artifact of the limiting procedure $X_0\rightarrow \infty$.

In Schwinger-Dyson equations the Keldysh component
of self-energy always appears between the powers of
retarded and advanced propagators. One easily finds that the mathematical
expression corresponding to such a product
is well defined even for multiple
self-energy insertion contributions.
We have studied single self-energy insertion in more detail.
We have obtained non-WTPF contribution
that generates nontrivial $X_0$ dependence.

In the case of production of photons from QCD plasma
(finite-lifetime effect), approximate analytic results
from our approach are almost identical
to those obtained by
S. -Y. Wang and D. Boyanovsky, who use the dynamical
renormalization group approach.

We may conclude that, indeed, out of equilibrium TFT,
using the finite-time path and the recognition of basic quantities
as WTPF's, retain all good properties of the Keldysh-time-path formulation
(energy-momentum space description, Feynman diagrams),
while removing the problem of illdefined quantities.

{\large \bf Acknowledgments}
I am especially indebted to Rolf Baier for
many stimulating discussions.
I also acknowledge
financial support from the A. von Humboldt foundation.
This work was supported by the Ministry of
Science and Technology
of the Republic of Croatia under contract No. 00980102.

{\bf Figure Captions}

Fig. 1: Finite switching-on time path.
\end{document}